# Direct demonstration of the emergent magnetism resulting from the multivalence Mn in a LaMnO$_3$ epitaxial thin film system

*Wei Niu, Wenqing Liu, Min Gu, Yongda Chen, Xiaoqian Zhang, Minhao Zhang, Yequan Chen, Ji Wang, Jun Du, Fengqi Song, Xiaoqing Pan, Nini Pryds, Xuefeng Wang,\* Peng Wang,\* Yongbing Xu,\* Yunzhong Chen,\* and Rong Zhang*

W. Niu, Y. D. Chen, X. Q. Zhang, M. H. Zhang, Y. Q. Chen, Prof. X. F. Wang, Prof. Y. B. Xu and Prof. R. Zhang
National Laboratory of Solid State Microstructures and Jiangsu Provincial Key Laboratory of Advanced Photonic and Electronic Materials, School of Electronic Science and Engineering, Nanjing University, Nanjing 210093, China
E-mail: xfwang@nju.edu.cn; ybxu@nju.edu.cn;

W. Niu, Prof. N. Pryds and Dr. Y. Z. Chen
Department of Energy Conversion and Storage, Technical University of Denmark, Risø Campus, Roskilde 4000, Denmark
E-mail: yunc@dtu.dk

Dr. W. Q. Liu
Department of Electronic Engineering, Royal Holloway, University of London, Egham, Surrey TW20 0EX, United Kingdom

M. Gu, Prof. P. Wang and Prof. X. Q. Pan
College of Engineering and Applied Sciences, Nanjing University, Nanjing 210093, China
E-mail: wangpeng@nju.edu.cn

J. Wang, Prof. J. Du and Prof. F. Q. Song
School of Physics, Nanjing University, Nanjing 210093, China

Prof. J. Du, Prof. F. Q. Song, Prof. X. Q. Pan, Prof. X. F. Wang, Prof. P. Wang, Prof. Y. B. Xu, and Prof. R. Zhang
Collaborative Innovation Center of Advanced Microstructures, Nanjing University, Nanjing 210093, China

Prof. X. Q. Pan
Department of Chemical Engineering and Materials Science, University of California Irvine, Irvine, CA 92697, USA

**Abstract:**

Atomically engineered oxide heterostructures provide a fertile ground for creating novel states. For example, a two-dimensional electron gas at the interface between two oxide insulators, giant thermoelectric Seebeck coefficient, emergent ferromagnetism from otherwise nonmagnetic components, and colossal ionic conductivity. Extensive research efforts reveal that oxygen deficiency or lattice strain play an important role in determining these unexpected properties. Herein, by studying the abrupt presence of robust ferromagnetism (up to 1.5 $\mu_B$/Mn) in $LaMnO_3$-based heterostructures, we find the multivalence states of Mn that play a decisive role in the emergence of ferromagnetism in the otherwise antiferromagnetic $LaMnO_3$ thin films. Combining spatially resolved electron energy-loss spectroscopy, X-ray absorption spectroscopy and X-ray magnetic circular dichroism techniques, we determine unambiguously that the ferromagnetism results from a conventional $Mn^{3+}$-O-$Mn^{4+}$ double-exchange mechanism rather than an interfacial effect. In contrast, the magnetic dead layer of 5 unit cell in proximity to the interface is found to be accompanied with the accumulation of $Mn^{2+}$ induced by electronic reconstruction. These findings provide a hitherto-unexplored multivalence state of Mn on the emergent magnetism in undoped manganite epitaxial thin films, such as $LaMnO_3$ and $BiMnO_3$, and shed new light on all-oxide spintronic devices.

# 1. Introduction

Complex oxides provide a rich platform for creating novel states and functional properties, especially at the interface of heterostructures and superlattices.[1] The interplay between charge, spin, orbital and lattice degrees of freedom gives rise to a rich spectrum of fascinating phenomena at complex oxide interfaces, including high-mobility two-dimensional electron gases,[2,3] giant thermoelectric Seebeck coefficient,[4] multiferroic properties,[5] colossal ionic conductivity[6] and intriguing magnetic properties.[7,8] Often, oxygen deficiency and/or strain effect play a key role in determining these unexpected properties.[6,9-12] For example, unintentional introduction of oxygen vacancies on the $SrTiO_3$ (STO) side contributes significantly to the metallic conduction at the interface of $LaAlO_3/SrTiO_3$ (LAO/STO).[13,14] The effect of oxygen excess at complex oxide interfaces has long been neglected since atomically controlled fabrication of oxide heterostructures are mostly achievable under oxygen deficient conditions. Consequently, to date, a comprehensive study to understand it and its role in the novel properties of the strongly correlated complex oxides has not been fully reached.

The stoichiometric bulk-state $LaMnO_3$ (LMO) is an A-type antiferromagnetic insulator with the orthorhombic perovskite crystal structure.[15,16] However, it turns to be a ferromagnetic insulator when formed as thin films.[17-19] The origin of such emergent ferromagnetic state has been vigorously investigated but the underlying mechanisms remains controversial.[17,18,20-22] Using scanning superconducting quantum interference device (SQUID) microscopy, ferromagnetism in LMO/STO heterostructure was attributed to stem from an electronic reconstruction at the interface due to the polar discontinuity.[17,18,23] However, such polarity-related electronic reconstruction normally occurs in proximity to the sample surface or interface. This can hardly explain the ferromagnetism which only occurs when the LMO film is thick. Besides, spin and orbital Ti magnetism was suspected to be responsible for the ferromagnetic behaviors in the LMO/STO superlattices.[21,24] Nevertheless, since most of the Mn $e_g$ bands locate approximately 1 eV below the conduction band of STO (Ti $3d$ $t_{2g}$), the Ti magnetism is expected to be quite small.[25,26] Moreover, the epitaxial strain induced by the lattice mismatch between the film and substrate has also been suggested an origins for the ferromagnetic state.[27,28] But, as we will see later, this can hardly explain that ferromagnetism is universally observed in LMO films grown on different substrates. In contrast, our demonstration of the multivalence state of Mn can explain both the previously reported data and the new results present here. Notably, oxygen excess is often induced by the cation vacancies[29-32] since the perovskite structure cannot accept the excessive oxygen in an interstitial site. Although oxygen excess has been put forward as a possible explanation for the ferromagnetism in very thick LMO films (100-150 nm),[33] it remains open whether such ferromagnetism can persist in proximity to the interface.

In this article, we report that the antiferromagnetic LMO can easily adopt the excessive oxygen from its stoichiometric phase, thus the multivalence states of Mn appear and show emergent ferromagnetism when epitaxially grown in thin film form via a well-controlled layer-by-layer two-dimensional growth mode. The ferromagnetism shows up abruptly when the film thickness is above 6 unit cell (uc). It is sensitive to the oxygen deposition pressure ($P_{O2}$) but insensitive to the type of the substrate (i.e., the strain). By combing spatially resolved electron energy-loss spectroscopy (EELS), element-specific X-ray absorption spectroscopy (XAS) and X-ray magnetic circular dichroism (XMCD), we find that the mixed valence state of Mn ions ($Mn^{4+}$, $Mn^{3+}$ and $Mn^{2+}$) exists and the ferromagnetism is attributed to the $Mn^{3+}$-O-$Mn^{4+}$ double-exchange mechanism. In addition, the magnetic dead-layer effect, i.e., the strong depression of magnetic properties when the thickness of the film is below 6 uc, is found to be associated with an accumulation of $Mn^{2+}$ induced by electronic reconstruction in the proximity of the interface.

## 2. Result and discussion

### 2.1. Emergent ferromagnetism in LMO-based heterostructures

LMO (001) thin films with thickness ranging from 3 to 20 uc were grown by pulsed laser deposition (PLD) on different perovskite substrates of STO (001), $(LaAlO_3)_{0.3}(Sr_2AlTaO_6)_{0.7}$ (LSAT) (001) and $LaAlO_3$ (LAO) (001), as schematically shown in **Figure 1**a. The film growth is in-situ monitored by the reflection high-energy electron diffraction (RHEED) under different $P_{O2}$ ($10^{-7}$-$10^{-3}$ mbar), where layer-by-layer film growth is achieved (Figure 1b). The sharp streaky line of the RHEED patterns in Figure 1c, d and e after deposition further indicate high quality of the films. All films show atomically smooth surface as confirmed by atomic force microscopy (Figure 1f). Note that, previous reports indicate that the as-deposited LMO thin films already show noticeable oxygen excess at a deposition pressure of $10^{-3}$ mbar.[33, 34]

**Figure 2**a shows the temperature-dependent magnetic moment (*M-T*) of LMO/STO heterostructures with different film thickness (*t*). When $t \leq 5$ uc, no ferromagnetism is observed. In contrast, at $t = 6$ uc, the LMO film abruptly turns ferromagnetic with Curie temperature ($T_C$) of ~95 K and a saturation magnetization of ~ 0.5 $\mu_B$/Mn (at *B* =0.1 T). Later on, the ferromagnetism develops gradually with the increase in the film thickness. Figure 2b summarizes the LMO saturation magnetic moment ($M_s$) as a function of thickness measured at 10 K. After the sharp onset of ferromagnetic order at $t = 6$ uc, the $M_s$ of the LMO/STO heterostructures increases with the increasing *t*. When $t \geq 10$ uc, the $M_s$ saturates at a value of ~1.5 $\mu_B$/Mn. This behavior indicates that the measured magnetism comes from the majority of the "bulk" of the film rather than an interfacial effect. If it were, $M_s$ would have shown a decrease as the thickness of the LMO layer increases.[21]

We further find that the highly ferromagnetic state is not limited to the LMO/STO heterostructures (the lattice parameter of STO is 3.91 Å), but can be universally observed even when LMO film is epitaxially grown on two other substrates with different lattice parameters and polarities, i.e. LSAT ($a$=3.86 Å) and LAO ($a$=3.79 Å). For comparison, Figure 2c shows the *M-T* curves for LMO films (20 uc) grown on STO, LSAT and LAO substrates, respectively. All these LMO-based heterostructures evidently demonstrate the emergent ferromagnetism despite the difference of $M_s$. The $M_s$ of heterostructures of LMO/STO, LMO/LSAT and LMO/LAO is found to be 1.56 $\mu_B$/Mn, 0.84 $\mu_B$/Mn and 0.72 $\mu_B$/Mn, respectively. The inset of Figure 2c displays the field-dependent in-plane magnetization ($T$ = 10 K) of these three heterostructures with dominant hysteresis loops, further suggesting the robust ferromagnetism. Since these epitaxial films exhibit different content of coherent strains,[18] the presence of ferromagnetism in these three different types of heterostructures clearly rules out the strain effect as the dominant origin for the occurrence of ferromagnetism.

To explore the influence of oxygen pressure on the magnetism, LMO films were further grown under different $P_{O2}$ conditions. Figure 2d shows a comparison of *M-T* curves of the LMO films ($t$ =10 uc) grown at different $P_{O2}$. With the $P_{O2}$ increasing from $10^{-7}$ mbar to $10^{-3}$ mbar, the magnetization is enhanced by nearly 6 times in magnitude up to a value of ~1.3 $\mu_B$/Mn. Moreover, the $T_C$ is also significantly increased from ~50 K to ~137 K upon the increase in $P_{O2}$ (inset of Figure 2d). Therefore, it is clear from these results that the ferromagnetism of LMO-based heterostructures shows strong dependence on the oxygen atmosphere during the deposition.

## 2.2. Multivalence state of Mn ions—STEM-EELS observation

**Figures 3**a and b show chemically atomic-resolution high-angle annular dark field (HAADF) scanning transmission electron microscope (STEM) images of the cross-section 10-uc LMO/STO heterostructure of samples grown at $1\times10^{-3}$ mbar and $1\times10^{-7}$ mbar, respectively. Both samples show a nice continuity of the perovskite-structure stacking

sequence across the interface. The interfacial layers are well crystallized without appreciable dislocations or other defects. This confirms the high-quality coherent layer-by-layer epitaxial growth. Notably, these two LMO samples exhibit exactly the same lattice parameters ($a=b=\sim3.884$ Å, $c=\sim3.914$ Å), i.e. the same interfacial strain state, but they show the different magnetic properties. This further excludes the possibility of lattice strain as a dominant origin of the underlying ferromagnetism. To reveal the intrinsic difference between these two samples, spatially resolved EELS[35] across the LMO/STO interface and on the LMO side is further measured. Figure 3d and e show the Mn-$L_{2,3}$ edge EELS mapping of the selected area in Figure 3a and b (red rectangles). It can be clearly seen that the Mn-$L_{2,3}$ edge appears two peaks, corresponding to the excitations from the spin-orbital splitting of $2p_{3/2}$ and $2p_{1/2}$ levels to empty states in the $3d$ band.[19]

Figure 3f shows the corresponding EELS profile of the ferromagnetic LMO sample (grown at $P_{O2}=1\times10^{-3}$ mbar) from the LMO/STO interface to the LMO film surface. Notably, Mn-$L_3$ edge spectra close to the interface exhibit a slight shift towards the lower energy loss in comparison to those from the surface of the LMO films, which suggests that the chemical valence state of Mn near the interface is reduced. A fit to the Mn-$L_3$ edge reveals the presence of multivalence state of $Mn^{2+}$, $Mn^{3+}$ and $Mn^{4+}$ (see Supporting Information Figure S3). The fitting results reveal that a fraction of $Mn^{2+}$ dominates at the interface (~42%) and then drops to a constant ratio (~20%) at 5 uc far away from the interface. However, the proportion of $Mn^{3+}/Mn^{4+}$ increases firstly to a maximum at 5 uc and then remains unchanged along with the increased thickness of LMO from the interface to the surface. These findings are further confirmed by the following XAS measurements. Notably, this trend of a reduction of Mn oxidation valence state is similar to the case of optimally doped manganite, such as the ferromagnetic $La_{0.7}Sr_{0.3}MnO_3$/STO[36-39] and the ferromagnetic $La_{0.7}Ca_{0.3}MnO_3$/STO.[40] In these doped manganite films, the ferromagnetism results from the $Mn^{3+}$-O-$Mn^{4+}$ double-exchange interaction. The reduced oxidation state ($Mn^{2+}$) in the sample corresponds to the decrease in the magnetization and has been proposed to explain the observed magnetic dead layer in the most manganite thin films.[38]

Figure 3c shows the comparison of O-$K$ edges of two samples. The relative strength of the first peak in green shaded area of O-$K$ edge is an indication of $d$-band filling.[41] Remarkably, the intensity of the first peak in O-$K$ spectra of LMO film grown at $1\times10^{-7}$ mbar is lower than the one grown at $1\times10^{-3}$ mbar. This reduction in the first-peak intensity of O-$K$ edge indicates that more electrons fill in the $d$-band. Therefore, the weak ferromagnetic LMO film grown at $1\times10^{-7}$ mbar has a lower valence state of Mn as compared to the robust ferromagnetic LMO film grown under $1\times10^{-3}$ mbar. The chemical valence state of Mn in the sample grown at $1\times10^{-7}$ mbar is dominated by $Mn^{3+}$ with a lower ratio of the $Mn^{4+}$. As shown in Figure 3g, we compare the Mn-$L_{2,3}$ edge spectra of 10-uc LMO grown under the different $P_{O2}$. It is obvious that there is a peak shift towards the higher energy loss for the LMO film grown under $1\times10^{-3}$ mbar, further proving evidence of the higher content of $Mn^{4+}$ under the high-oxygen-pressure condition. More quantitative analysis reveals that, compared with the $Mn^{4+}$ ratio of ~24.6% in the $1\times10^{-3}$ mbar sample, the ratio of $Mn^{4+}$ decreases to ~15.9% in the $1\times10^{-7}$ mbar sample, while the proportion of $Mn^{3+}$ remains nearly unchanged. In addition, the portion of $Mn^{2+}$ increases from 19.6% to 24.5% as the $P_{O2}$ decreases. From these results, we draw the conclusion that the main difference between these two LMO films is that, the ferromagnetic LMO film grown under higher $P_{O2}$ has more $Mn^{4+}$ ions and thus a high $Mn^{4+}/Mn^{3+}$ ratio, which enhances the double-exchange interaction between $Mn^{3+}$ and $Mn^{4+}$.

**2.3. Multivalence state of Mn ions—XAS/XMCD observation**

To further unveil the origin of the emergent ferromagnetism, the element-specific XAS and XMCD at the Mn and Ti $L_{2,3}$ absorption edges are performed to probe the local electronic character of the magnetic ground state of the LMO. As schematically shown in **Figure 4**a, circularly polarized X-rays with ~100% degree of polarization were performed at 60º with respect to the film plane and in parallel with the applied magnetic field. Typical XAS spectra of the LMO/STO heterostructures are displayed in Figure 4b. Mn-$L_{2,3}$ edge XAS spectra show the prominent multiplet structure for both spin-orbit split core levels, indicating a mixture valence state of Mn. For comparison, spatially resolved XAS spectra and the corresponding atomic multiplet calculation of $Mn^{2+}$, $Mn^{3+}$ and $Mn^{4+}$ from Ref. [42] are taken as a reference (see the blue curves in Figure 4b). Remarkably, the experimental spectra are comparable to those reference spectra, showing that the experimental spectra are composed of the $Mn^{2+}$, $Mn^{3+}$ and $Mn^{4+}$ in all LMO films, in good agreement with our EELS results. $L_3$ peak locates at ~642.0 eV, indicating the main valence state of $Mn^{3+}$. While the stoichiometric LMO crystal contains only $Mn^{3+}$ ions, oxygen excess introduces $Mn^{4+}$ ions into the Mn sublattice, resulting in the observed ferromagnetic coupling between the local spins via double-exchange mechanism.[19,43] The robust ferromagnetism is favored in a more oxidizing atmosphere because of the formation of $Mn^{4+}$ and the resultant double-exchange mechanism relevant to $Mn^{3+}$-O-$Mn^{4+}$. The fingerprint of $Mn^{2+}$ valence state at the lower energies of ~640.1 eV dominates when $t < 6$ uc, while the $Mn^{3+}$ and $Mn^{4+}$ content increase with the increasing $t$ as evidenced by the increase in spectral weight at the photon energy of 642.0 eV and 643.4 eV. Note that the increased $Mn^{4+}/Mn^{3+}$ ratio with the thickness can rigorously increase the strength of the double-exchange interaction, thereby enhancing the ferromagnetism. Moreover, the spectra exhibit the identical line-shapes, indicating that LMO has the same La/Mn ratio, otherwise a different ratio should have resulted in a different line-shape.[44] This is also consistent with the bulk LMO case. Although an oxygen excess can be obtained, the La/Mn ratio always keeps 1.[31]

Ti-$L_{2,3}$ isotropic XAS measurements for heterostructures with different LMO thickness are shown in Figure 4c. All measurements display a similar electronic structure. The energy difference between the two main peaks of the $L_3$ and $L_2$ edges remain unchanged, as expected for $Ti^{4+}$ of the STO layer. This suggests no charge transfer to the Ti site. And the charge reconstruction occurs mainly on the Mn site at the interface of LMO/STO. The presence of $Ti^{4+}$ rather than $Ti^{3+}$ is also proved independently by our EELS spectra at Ti $L$-edge (Figure S3), consistent with the previous results,[37] ruling out the possible mechanism of Ti interfacial magnetism.[21,24]

In the light of the valence state of Mn ions, we now turn to its contribution to magnetic properties. Representative XAS/XMCD and their integration taken at the Mn-$L_{2,3}$ edges are shown in **Figure 5**a. The magnitude of the magnetization is quantitatively estimated using the XMCD spin sum-rules.[45] Within the limit of the uncertainties in the sum-rules estimation, the obtained total magnetization is approximately ~1.5 $\mu_B$/Mn, which is in reasonable agreement with our SQUID results. In addition, the XMCD signal of ~24% at the Mn-$L_3$ edge of 15-uc LMO is consistent with previously measured values.[45] We summarize the thickness dependence of the XMCD-derived magnetic moment in Figure 5b, together with the thickness-dependent magnetization of LMO/STO measured by SQUID. Good agreement of the magnetization derived from both measurements show the same thickness dependence. These results indicate that the ferromagnetism of LMO/STO heterostructures comes from the intrinsic contribution from the exchange interaction of the multivalence Mn ions rather than the interface-driven effect from the electronic reconstruction or the strain-induced effect by the substrates.

**2.4. Electronic reconstruction and dead-layer behavior in manganite films**

For bulk LMO, oxygen excess is usually accommodated by the formation of cation vacancies on both La and Mn sites,[31,46] leaving a perfect oxygen sublattice. To keep neutrality of the charge, a fraction of $Mn^{3+}$ in the stoichiometric LMO must be oxidized to $Mn^{4+}$. This is a direct reason that the mixed valence states of Mn ($Mn^{3+}$ and $Mn^{4+}$) should be shown in the EELS and XAS spectra. The oxygen excess harasses ferromagnetic $Mn^{3+}$-O-$Mn^{4+}$ double-exchange interactions by the $Mn^{4+}/Mn^{3+}$ ratio.[31,32] The ferromagnetic $BiMnO_3$ holds the similar mechanism to LMO due to the oxygen excess.[47] Based on our comparison of O-$K$ and Mn-$L$ edges of LMO deposited at the different oxygen pressure (Figure 3), when the LMO film is grown at $P_{O2}=1\times10^{-7}$ mbar the ratio of $Mn^{4+}/Mn^{3+}$ is reduced, thereby the double-exchange interaction is weakened and the ferromagnetism is depressed.[48] In contrast, when deposited at a typical oxygen pressure of $10^{-3}$ mbar, the higher the $Mn^{4+}/Mn^{3+}$ ratio the more significant is the double-exchange interaction as well as the robust ferromagnetism.

Additionally, LMO is polar whereas STO is nonpolar. Therefore, polar-discontinuity-induced electronic reconstruction could occur at the LMO/STO interface.[2] However, different from the intensively investigated LAO/STO system where the reconstructed electrons are transferred from the sample surface to the STO inside, in LMO/STO system, the empty or partially filled $e_g$ bands of LMO are often lower than the Ti $3d$ bands of STO,[49] as schematically illustrated in **Figure 6**a. In this vein, the electronic reconstruction in LMO/STO occurs on the LMO side, i.e. from LMO surface to the LMO in proximity to the interface rather than to the STO side.[49-51] This scenario is consistent with the fact that $Mn^{2+}$ becomes detectable at the interface while no signal of $Ti^{3+}$ is detected, as evidenced in the EELS and XAS spectra of Ti (Figure 4 and Figure S3). It should be noted that,[18] Wang *et al.* did not observe ferromagnetism when LMO thin films (12 uc) were grown on the polar substrates of LAO and LSAT. They attributed the non-ferromagnetism to the absence of polar discontinuity because both LMO and substrates (LAO and LSAT) are polar.[18] Herein, when we increase the film thickness to 20 uc, all heterostructures of LMO/STO, LMO/LAO and LMO/LSAT show ferromagnetism (Figure 1c). Hence, the electronic reconstruction due to the polar discontinuity cannot account for the ferromagnetism of LMO but coherent with the absence of magnetism at the interface.

The sharp transition from ferromagnetism to non-ferromagnetism at $t < 6$ uc should suggest a magnetic dead layer of LMO in the vicinity of the interface. As shown in Figure 6b, LMO thin films are spatially separated into two stacked sheets. One is dead layer at the bottom without ferromagnetism, the other is a uniformly magnetized layer on the top. The dead layer is widely observed in other ferromagnetic manganite films, such as $La_{1-x}Sr_xMnO_3$ ($x$~0.3 and 0.33) and LMO, as summarized in Figure 6c.[52-54] The mechanism underlying this dead-layer effect remains elusive. However, this might be due to the presence of $Mn^{2+}$,[38] which always shows a critical thickness ($t_c$) of ~2 nm (i.e., 5 uc) when the substrate is STO. The different critical thickness of the dead layer on various substrates could be due to the fact that films on the different substrates possess the different responses, such as the energy cost of creating vacancies and the capacity of adapting chemical stoichiometry during the deposition.[55-57]

## 3. Conclusion

To conclude, we have firstly established the multivalence state of $Mn^{2+}$, $Mn^{3+}$ and $Mn^{4+}$ ions in LMO epitaxial thin films. The ferromagnetism originates from a conventional $Mn^{3+}$-O-$Mn^{4+}$ double-exchange mechanism rather than an interface or strain effect. The observed thickness-dependent ferromagnetism is controlled by the magnetic dead-layer effect in manganite thin films, which is accompanied by the accumulation of $Mn^{2+}$ induced by the electronic reconstruction. Our results not only shed light on the emergent ferromagnetism of the otherwise antiferromagnetic LMO thin films, but also broaden the understanding of oxygen excess in complex oxides, which is beneficial for the rational design of future all-oxide spintronic devices, particularly based on LMO.

## 4. Experimental Section

*Epitaxial Growth*: The LMO films were grown by PLD technique (KrF excimer laser with $\lambda$=248 nm and $E$=100 mJ) as described elsewhere.[58] The growth dynamics was investigated by monitoring the RHEED pattern. During the growth, the temperature was maintained at 750ºC. Without particularly mentioned, all LMO films were grown in an oxygen environment at a pressure of 1×10$^{-3}$ mbar. After the deposition, the samples were slowly cooled down to room temperature at the growth pressure without further post-annealing.

*STEM-EELS Measurements*: Microstructures of cross-sectional samples were investigated using an aberration-corrected STEM Titan at 300 keV. Elemental analysis was carried out using STEM-EELS spectrum imaging with a Gatan Quantum 966 system. The near-edge fine structures were used to study the local electronic structures at a sub-uc level of resolution with an energy resolution of 0.1 eV.

*Magnetic Measurements*: A Quantum Design SQUID measurement system was used to measure the magnetic properties from 10 to 300 K with the magnetic field applied in-plane along the (100) direction of the substrate. The temperature-dependent magnetic measurements were performed by first cooling in 1 T, and the in-plane magnetic moments were then measured during warm-up in 0.1 T.[18]

*X-Ray Absorption Measurements*: XAS and XMCD at the Ti and Mn $L_{2,3}$ absorption edges were performed on Beamline I06 at the Diamond Light Source, UK. Oppositely circular polarized X-rays with 100% polarization degree were used successively to resolve XMCD signals from Mn and Ti. The light-helicity was switched in a saturating magnetic field of 1 T, which was applied at 60° with respect to the film plane and in parallel with the incident beam. XAS was obtained in total electron yield (TEY) mode at 10 K. The probing thickness was ~3-5 nm. XMCD was calculated as $|\mu^+ - \mu^-|/(\mu^+ + \mu^-)$, where $\mu$ represents the TEY-XAS intensity for the respective helicities of the emitted light.

**Supporting Information**
Supporting Information is available from the Wiley Online Library or from the author.

**Acknowledgements**

W. Niu, W. Q. Liu, M. Gu, and Y. D. Chen contributed equally to this work. X. F. Wang and Y. Z. Chen conceived the study and designed the experiments. W. Niu and Y. D. Chen prepared and characterized the samples. M. Gu and P. Wang performed the STEM/EELS measurements. W. Q. Liu, X. Q. Zhang, and Y. B. Xu conducted the synchrotron measurements. M. H. Zhang and Y. Q. Chen performed the partial control experiments. J. Wang and J. Du carried out the SQUID measurements. F. Q. Song, X. Q. Pan, N. Pryds, and R. Zhang contributed to the data analysis. W. Niu, X. F. Wang and Y. Z. Chen wrote the paper. This work was financially supported by the National Key Basic Research Program of China under Grant Nos. 2014CB921103, 2015CB654901 and 2017YFA0206304, the National Natural Science Foundation of China under Grant Nos. U1732159, 11274003, 11474147, 91421109, 11522432, 61427812 and 11574288, the Priority Academic Program Development of Jiangsu Higher Education Institutions, and Collaborative Innovation Center of Solid-State Lighting and Energy-Saving Electronics.

**References**

[1]   H. Y. Hwang, Y. Iwasa, M. Kawasaki, B. Keimer, N. Nagaosa, Y. Tokura, *Nat. Mater.* **2012**, 11, 103.
[2]   A. Ohtomo, H. Y. Hwang, *Nature* **2004**, 427, 423.
[3]   Y. Z. Chen, N. Bovet, F. Trier, D. V. Christensen, F. M. Qu, N. H. Andersen, T. Kasama, W. Zhang, R. Giraud, J. Dufouleur, T. S. Jespersen, J. R. Sun, A. Smith, J. Nygard, L. Lu, B. Buchner, B. G. Shen, S. Linderoth, N. Pryds, *Nat. Commun.* **2013**, 4, 1371.
[4]   H. Ohta, S. Kim, Y. Mune, T. Mizoguchi, K. Nomura, S. Ohta, T. Nomura, Y. Nakanishi, Y. Ikuhara, M. Hirano, H. Hosono, K. Koumoto, *Nat. Mater.* **2007**, 6, 129.
[5]   A. R. Damodaran, J. D. Clarkson, Z. Hong, H. Liu, A. K. Yadav, C. T. Nelson, S. L. Hsu, M. R. McCarter, K. D. Park, V. Kravtsov, A. Farhan, Y. Dong, Z. Cai, H. Zhou, P. Aguado-Puente, P. Garcia-Fernandez, J. Iniguez, J. Junquera, A. Scholl, M. B. Raschke, L. Q. Chen, D. D. Fong, R. Ramesh, L. W. Martin, *Nat. Mater.* **2017**, 16, 1003.
[6]   J. A. Kilner, *Nat. Mater.* **2008**, 7, 838.
[7]   B. Chen, H. Xu, C. Ma, S. Mattauch, D. Lan, F. Jin, Z. Guo, S. Wan, P. Chen, G. Gao, F. Chen, Y. Su, W. Wu, *Science* **2017**, 357, 191.
[8]   D. Pesquera, G. Herranz, A. Barla, E. Pellegrin, F. Bondino, E. Magnano, F. Sanchez, J. Fontcuberta, *Nat. Commun.* **2012**, 3, 1189.
[9]   J. N. Eckstein, *Nat. Mater.* **2007**, 6, 473.
[10]  M. Huijben, G. Koster, M. K. Kruize, S. Wenderich, J. Verbeeck, S. Bals, E. Slooten, B. Shi, H. J. A. Molegraaf, J. E. Kleibeuker, S. van Aert, J. B. Goedkoop, A. Brinkman, D. H. A. Blank, M. S. Golden, G. van Tendeloo, H. Hilgenkamp, G. Rijnders, *Adv. Func. Mater.* **2013**, 23, 5240.
[11]  S. Cheng, M. Li, S. Deng, S. Bao, P. Tang, W. Duan, J. Ma, C. Nan, J. Zhu, *Adv. Funct. Mater.* **2016**, 26, 3589.
[12]  H.-J. Liu, T.-C. Wei, Y.-M. Zhu, R.-R. Liu, W.-Y. Tzeng, C.-Y. Tsai, Q. Zhan, C.-W. Luo, P. Yu, J.-H. He, Y.-H. Chu, Q. He, *Advanced Functional Materials* **2016**, 26, 729.
[13]  G. Herranz, M. Basletic, M. Bibes, C. Carretero, E. Tafra, E. Jacquet, K. Bouzehouane, C. Deranlot, A. Hamzic, J. M. Broto, A. Barthelemy, A. Fert, *Phys. Rev. Lett.* **2007**, 98, 216803.
[14]  Y. Chen, N. Pryds, J. E. Kleibeuker, G. Koster, J. Sun, E. Stamate, B. Shen, G. Rijnders, S. Linderoth, *Nano Lett.* **2011**, 11, 3774.
[15]  M. B. Salamon, M. Jaime, *Rev. Mod. Phys.* **2001**, 73, 583.
[16]  D. Yi, N. Lu, X. Chen, S. Shen, P. Yu, *J. Phys.: Condens. Matter* **2017**, 29, 443004.
[17]  Y. Anahory, L. Embon, C. J. Li, S. Banerjee, A. Meltzer, H. R. Naren, A. Yakovenko, J. Cuppens, Y. Myasoedov, M. L. Rappaport, M. E. Huber, K. Michaeli, T. Venkatesan, Ariando, E. Zeldov, *Nat. Commun.* **2016**, 7, 12566.

[18]     X. R. Wang, C. J. Li, W. M. Lü, T. R. Paudel, D. P. Leusink, M. Hoek, N. Poccia, A. Vailionis, T. Venkatesan, J. M. D. Coey, E. Y. Tsymbal, Ariando, H. Hilgenkamp, *Science* **2015**, 349, 716.
[19]     J. J. Peng, C. Song, B. Cui, F. Li, H. J. Mao, Y. Y. Wang, G. Y. Wang, F. Pan, *Phys. Rev. B* **2014**, 89, 165129.
[20]     X. Zhai, L. Cheng, Y. Liu, C. M. Schleputz, S. Dong, H. Li, X. Zhang, S. Chu, L. Zheng, J. Zhang, A. Zhao, H. Hong, A. Bhattacharya, J. N. Eckstein, C. Zeng, *Nat. Commun.* **2014**, 5, 4283.
[21]     J. Garcia-Barriocanal, J. C. Cezar, F. Y. Bruno, P. Thakur, N. B. Brookes, C. Utfeld, A. Rivera-Calzada, S. R. Giblin, J. W. Taylor, J. A. Duffy, S. B. Dugdale, T. Nakamura, K. Kodama, C. Leon, S. Okamoto, J. Santamaria, *Nat. Commun.* **2010**, 1, 82.
[22]     F. Hellman, A. Hoffmann, Y. Tserkovnyak, G. S. D. Beach, E. E. Fullerton, C. Leighton, A. H. MacDonald, D. C. Ralph, D. A. Arena, H. A. Dürr, P. Fischer, J. Grollier, J. P. Heremans, T. Jungwirth, A. V. Kimel, B. Koopmans, I. N. Krivorotov, S. J. May, A. K. Petford-Long, J. M. Rondinelli, N. Samarth, I. K. Schuller, A. N. Slavin, M. D. Stiles, O. Tchernyshyov, A. Thiaville, B. L. Zink, *Rev. Mod. Phys.* **2017**, 89, 025006.
[23]     Z. Chen, Z. Chen, Z. Q. Liu, M. E. Holtz, C. J. Li, X. R. Wang, W. M. Lü, M. Motapothula, L. S. Fan, J. A. Turcaud, L. R. Dedon, C. Frederick, R. J. Xu, R. Gao, A. T. N'Diaye, E. Arenholz, J. A. Mundy, T. Venkatesan, D. A. Muller, L.-W. Wang, J. Liu, L. W. Martin, *Phys. Rev. Lett.* **2017**, 119, 156801.
[24]     J. Garcia-Barriocanal, F. Y. Bruno, A. Rivera-Calzada, Z. Sefrioui, N. M. Nemes, M. Garcia-Hernandez, J. Rubio-Zuazo, G. R. Castro, M. Varela, S. J. Pennycook, C. Leon, J. Santamaria, *Adv Mater* **2010**, 22, 627.
[25]     J. Matsuno, A. Sawa, M. Kawasaki, Y. Tokura, *Appl. Phys. Lett.* **2008**, 92, 122104.
[26]     Y. Z. Chen, J. R. Sun, A. D. Wei, W. M. Lu, S. Liang, B. G. Shen, *Appl. Phys. Lett.* **2008**, 93, 152515.
[27]     M. An, Y. Weng, H. Zhang, J.-J. Zhang, Y. Zhang, S. Dong, *Phys. Rev. B* **2017**, 96, 235112.
[28]     J. Ma, Y. Zhang, L. Wu, C. Song, Q. Zhang, J. Zhang, J. Ma, C.-W. Nan, *MRS Commun.* **2016**, 6, 354.
[29]     J. A. M. V. Roosmalen, E. H. P. Cordfunke, *J. Solid State Chem.* **1994**, 110, 109.
[30]     A. Arulraj, R. Mahesh, G. N. Subbanna, R. Mahendiran, A. K. Raychaudhuri, C. N. R. Rao, *J. Solid State Chem.* **1996**, 127, 87.
[31]     J. Töpfer, J. B. Goodenough, *J. Solid State Chem.* **1997**, 130, 117.
[32]     C. Ritter, M. R. Ibarra, J. M. D. Teresa, P. A. Algarabel, C. Marquina, J. Blasco, J. García, S. Oseroff, S.-W. Cheong, *Phys. Rev.B* **1997**, 56, 8902.
[33]     Z. Marton, S. S. A. Seo, T. Egami, H. N. Lee, *J. Cryst. Growth* **2010**, 312, 2923.
[34]     H. S. Kim, H. M. Christen, *J. Phys.: Condens. Matter* **2010**, 22, 146007.
[35]     X. Wang, F. Song, Q. Chen, T. Wang, J. Wang, P. Liu, M. Shen, J. Wan, G. Wang, J.-B. Xu, *J. Am. Chem. Soc.* **2010**, 132, 6492.
[36]     Z. Yuan, J. Ruan, L. Xie, X. Pan, D. Wu, P. Wang, *Appl. Phys. Lett.* **2017**, 110, 171602.
[37]     J. A. Mundy, Y. Hikita, T. Hidaka, T. Yajima, T. Higuchi, H. Y. Hwang, D. A. Muller, L. F. Kourkoutis, *Nat. Commun.* **2014**, 5, 3464.
[38]     M. Nord, P. E. Vullum, M. Moreau, J. E. Boschker, S. M. Selbach, R. Holmestad, T. Tybell, *Appl. Phys. Lett.* **2015**, 106, 041604.
[39]     B. Cui, C. Song, G. A. Gehring, F. Li, G. Wang, C. Chen, J. Peng, H. Mao, F. Zeng, F. Pan, *Advanced Functional Materials* **2015**, 25, 864.
[40]     A. L. Kobrinskii, A. M. Goldman, M. Varela, S. J. Pennycook, *Phys. Rev. B* **2009**, 79, 094405.

[41] A. B. Shah, Q. M. Ramasse, X. Zhai, J. G. Wen, S. J. May, I. Petrov, A. Bhattacharya, P. Abbamonte, J. N. Eckstein, J.-M. Zuo, *Adv. Mater.* **2010**, 22, 1156.
[42] T. Burnus, Z. Hu, H. H. Hsieh, V. L. J. Joly, P. A. Joy, M. W. Haverkort, H. Wu, A. Tanaka, H. J. Lin, C. T. Chen, L. H. Tjeng, *Phys. Rev. B* **2008**, 77, 125124.
[43] C. Zener, *Phys. Rev.* **1951**, 81, 440.
[44] A. Galdi, C. Aruta, P. Orgiani, N. B. Brookes, G. Ghiringhelli, M. Moretti Sala, R. V. K. Mangalam, W. Prellier, U. Lüders, L. Maritato, *Phys. Rev. B* **2011**, 83, 064418.
[45] P. Yu, J. S. Lee, S. Okamoto, M. D. Rossell, M. Huijben, C. H. Yang, Q. He, J. X. Zhang, S. Y. Yang, M. J. Lee, Q. M. Ramasse, R. Erni, Y. H. Chu, D. A. Arena, C. C. Kao, L. W. Martin, R. Ramesh, *Phys. Rev. Lett.* **2010**, 105, 027201.
[46] Z. Li, M. Bosman, Z. Yang, P. Ren, L. Wang, L. Cao, X. Yu, C. Ke, M. B. H. Breese, A. Rusydi, W. Zhu, Z. Dong, Y. L. Foo, *Adv. Funct. Mater.* **2012**, 22, 4312.
[47] A. A. Belik, K. Kodama, N. Igawa, S.-i. Shamoto, K. Kosuda, E. Takayama-Muromachi, *J. Am. Chem. Soc.* **2010**, 132.
[48] L. Jin, C.-L. Jia, I. Lindfors-Vrejoiu, X. Zhong, H. Du, R. E. Dunin-Borkowski, *Adv. Mater. Interfaces* **2016**, 3, 1600414.
[49] Y. Z. Chen, F. Trier, T. Wijnands, R. J. Green, N. Gauquelin, R. Egoavil, D. V. Christensen, G. Koster, M. Huijben, N. Bovet, S. Macke, F. He, R. Sutarto, N. H. Andersen, J. A. Sulpizio, M. Honig, G. E. Prawiroatmodjo, T. S. Jespersen, S. Linderoth, S. Ilani, J. Verbeeck, G. Van Tendeloo, G. Rijnders, G. A. Sawatzky, N. Pryds, *Nat. Mater.* **2015**, 14, 801.
[50] J. J. Peng, C. Song, F. Li, Y. D. Gu, G. Y. Wang, F. Pan, *Phys. Rev. B* **2016**, 94, 214404.
[51] W. Niu, Y. Gan, Y. Zhang, D. V. Christensen, M. v. Soosten, X. Wang, Y. Xu, R. Zhang, N. Pryds, Y. Chen, *Appl. Phys. Lett.* **2017**, 111, 021602.
[52] R. P. Borges, W. Guichard, J. G. Lunney, J. M. D. Coey, F. Ott, *J. Appl. Phys.* **2001**, 89, 3868.
[53] M. Huijben, L. W. Martin, Y. H. Chu, M. B. Holcomb, P. Yu, G. Rijnders, D. H. A. Blank, R. Ramesh, *Phys. Rev. B* **2008**, 78, 094413.
[54] R. Peng, H. C. Xu, M. Xia, J. F. Zhao, X. Xie, D. F. Xu, B. P. Xie, D. L. Feng, *Appl. Phys. Lett.* **2014**, 104, 081606.
[55] S. Estradé, J. M. Rebled, J. Arbiol, F. Peiró, I. C. Infante, G. Herranz, F. Sánchez, J. Fontcuberta, R. Córdoba, B. G. Mendis, A. L. Bleloch, *Appl. Phys. Lett.* **2009**, 95, 072507.
[56] C. Cazorla, *Phys. Rev. Appl.* **2017**, 7, 044025.
[57] U. Aschauer, R. Pfenninger, S. M. Selbach, T. Grande, N. A. Spaldin, *Phys. Rev. B* **2013**, 88, 054111.
[58] W. Niu, M. Gao, X. Wang, F. Song, J. Du, X. Wang, Y. Xu, R. Zhang, *Sci. Rep.* **2016**, 6, 26081.

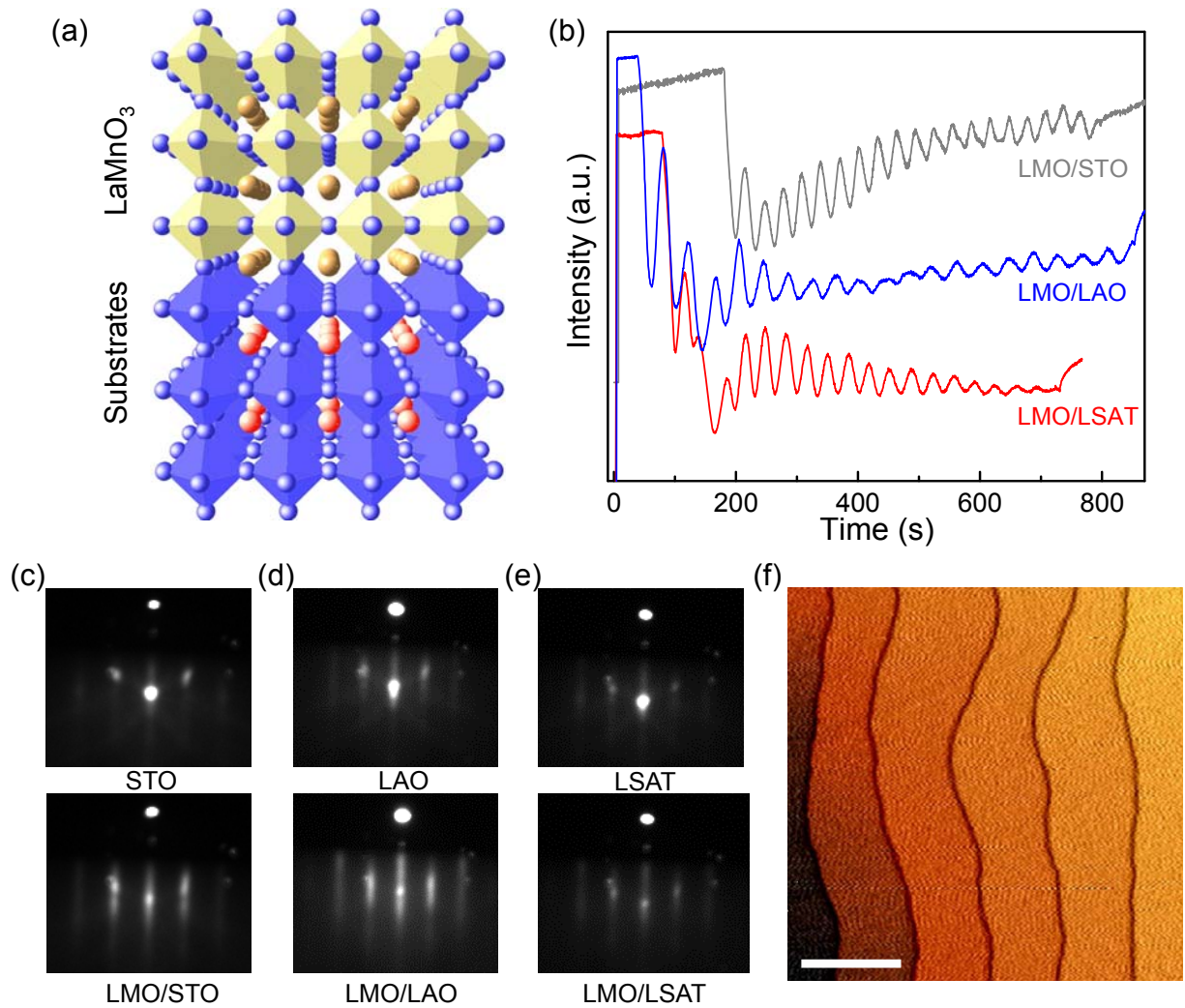

**Figure 1. Layer-by-layer epitaxial growth of LMO films on STO, LSAT and LAO substrates.** (a) Sketch of the LMO-based perovskite heterostructures. (b) Representative RHEED intensity oscillations for the typical 20-uc LMO film on various substrates. (c-e) RHEED patterns of STO, LAO and LSAT prior to the growth and 20-uc LMO films after the growth on these substrates. (f) Surface morphology (2 μm × 2 μm) of LMO films with thickness of 20 uc on STO. The scale bar is 500 nm.

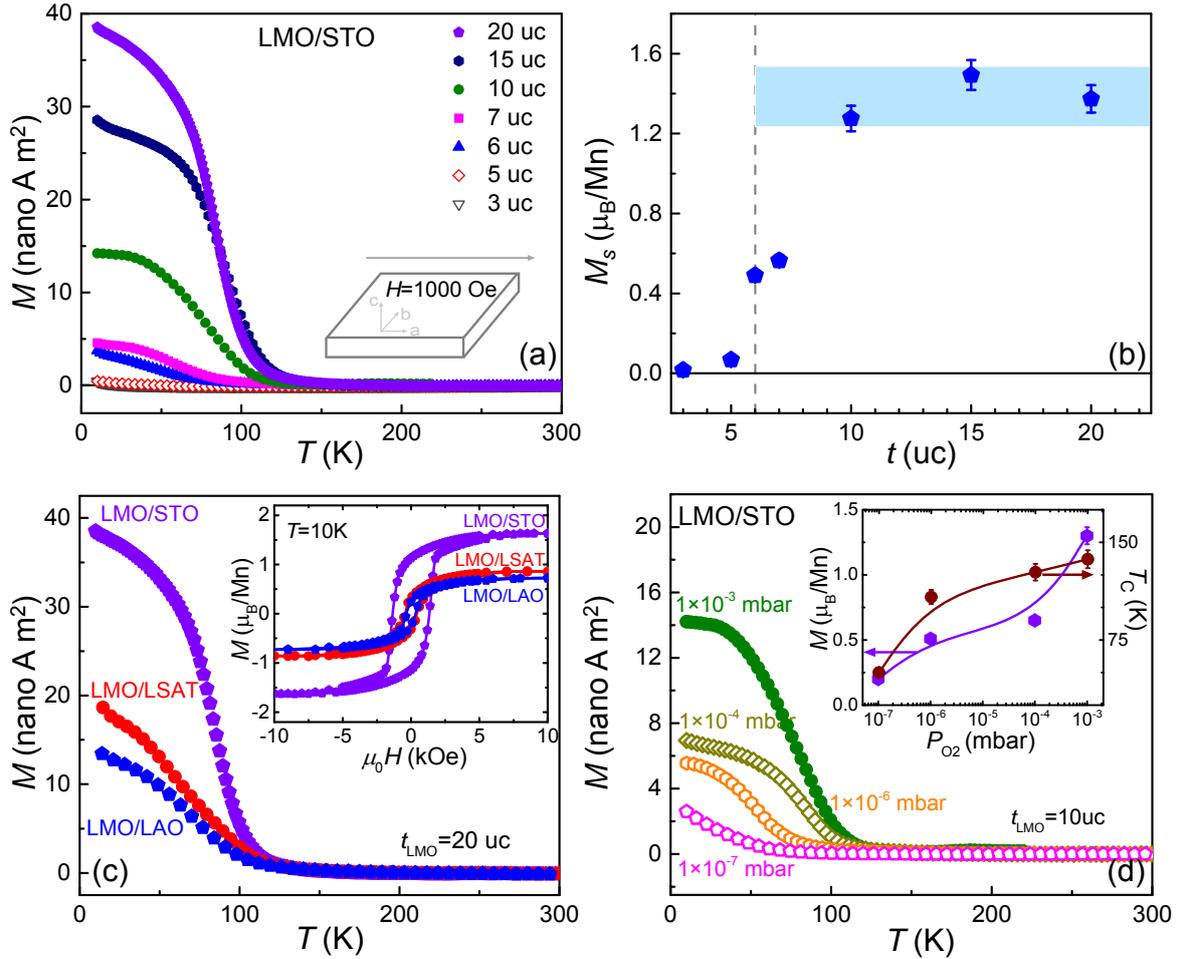

**Figure 2. Emergent ferromagnetism in LMO-based heterostructures.** (a) Temperature dependence of magnetic moments of LMO/STO heterostructures with the thickness ranging from 3 to 20 uc. The inset shows the magnetization measurement configuration. (b) Thickness-dependent saturation magnetic moment of LMO/STO heterostructure at 10 K. (c) Magnetic moment of 20-uc LMO films as a function of temperature grown on STO, LSAT and LAO substrates, respectively. The inset shows the *M-H* curves of LMO grown on these substrates. (d) Magnetic moment as a function of temperature of 10-uc LMO films grown under different $P_{O2}$. The inset shows the $M_s$ and $T_C$ as a function of the $P_{O2}$.

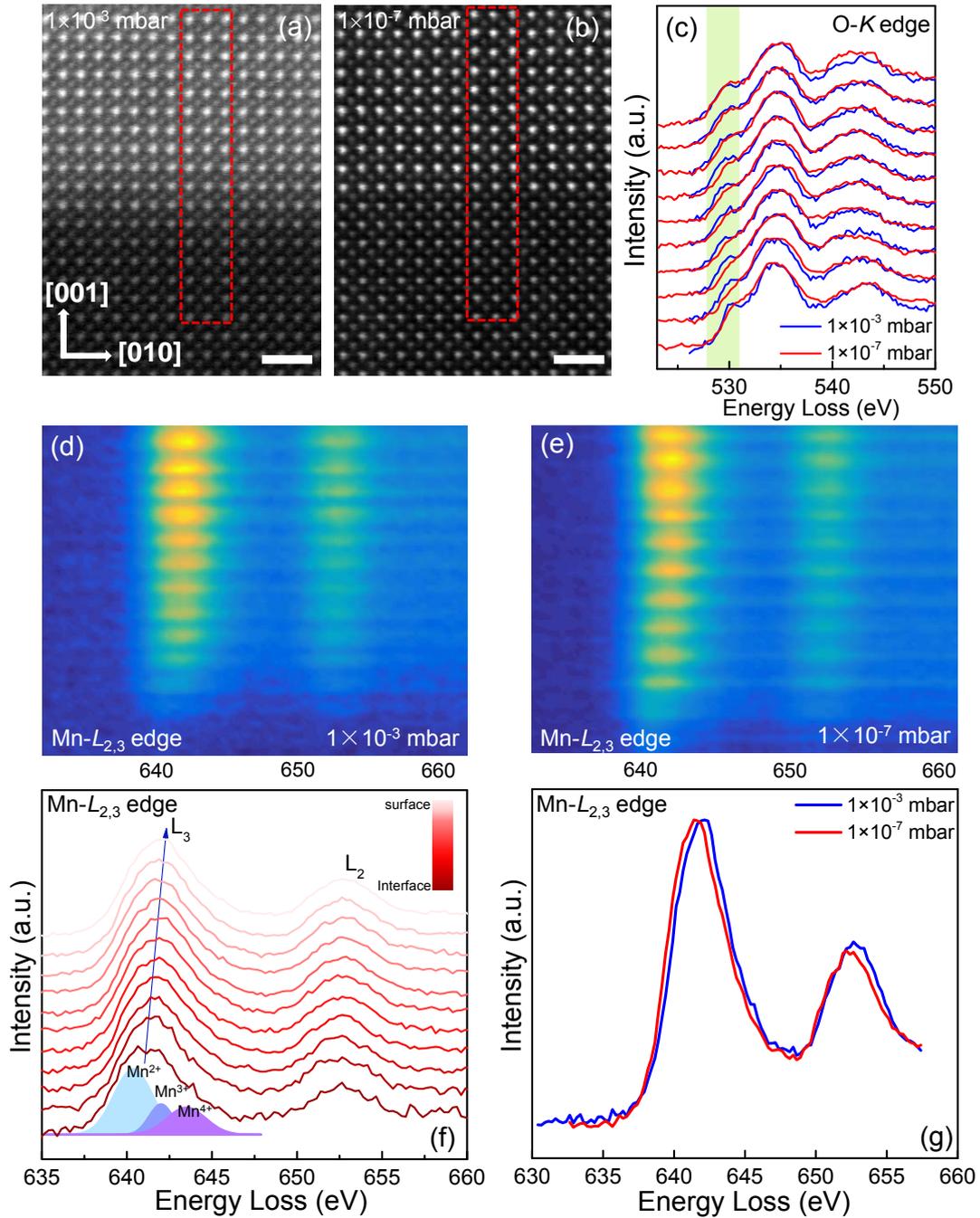

**Figure 3. Valence state variations of Mn ions at different $P_{O2}$.** (a) and (b) Cross-sectional STEM-HAADF images of 10 uc-LMO/STO grown at $1\times10^{-3}$ mbar and $1\times10^{-7}$ mbar, respectively. The scale bar is 1 nm. (c) The EELS profile comparison of O-$K$ edge of 10-uc LMO grown at the different oxygen pressure. (d) and (e) EELS mapping of Mn-$L_{2,3}$ edge of each layer in the selected interfacial area indicated in (a) and (b). (f) Corresponding EELS profiles of Mn-$L_{2,3}$ edge of (d). (g) Comparison of Mn-$L_{2,3}$ edge EELS spectra between 10-uc LMO films grown at the different oxygen pressure.

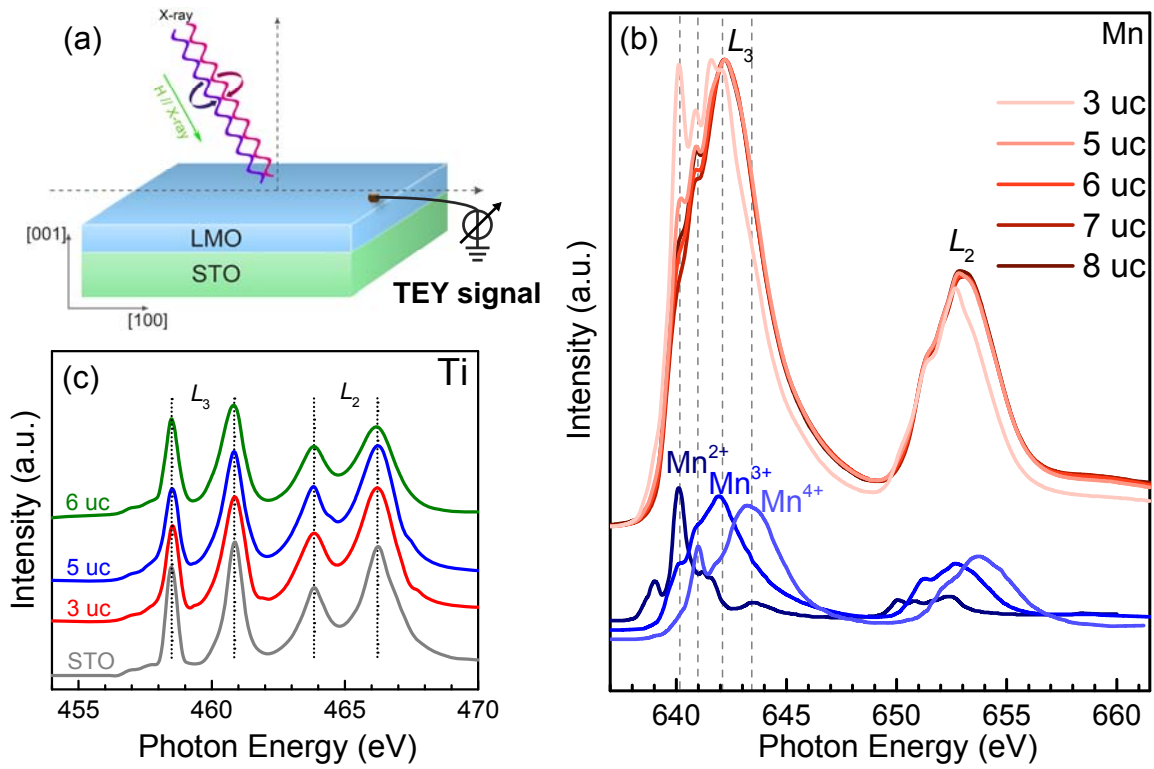

**Figure 4. Mn and Ti XAS spectra of LMO/STO heterostructures.** (a) The schematic diagram of the experimental configuration for the XAS/XMCD measurements. (b) Normalized isotropic XAS spectra at Mn-$L_{2,3}$ edges of LMO/STO heterostructures at 10 K and 1 T. The marked four dashed lines from left to right indicate the peak positions of $Mn^{2+}$, $Mn^{4+}$, $Mn^{3+}$ and $Mn^{4+}$, respectively. (c) Ti-$L_{2,3}$ edge XAS spectra of 3-uc, 5-uc and 6-uc LMO on STO.

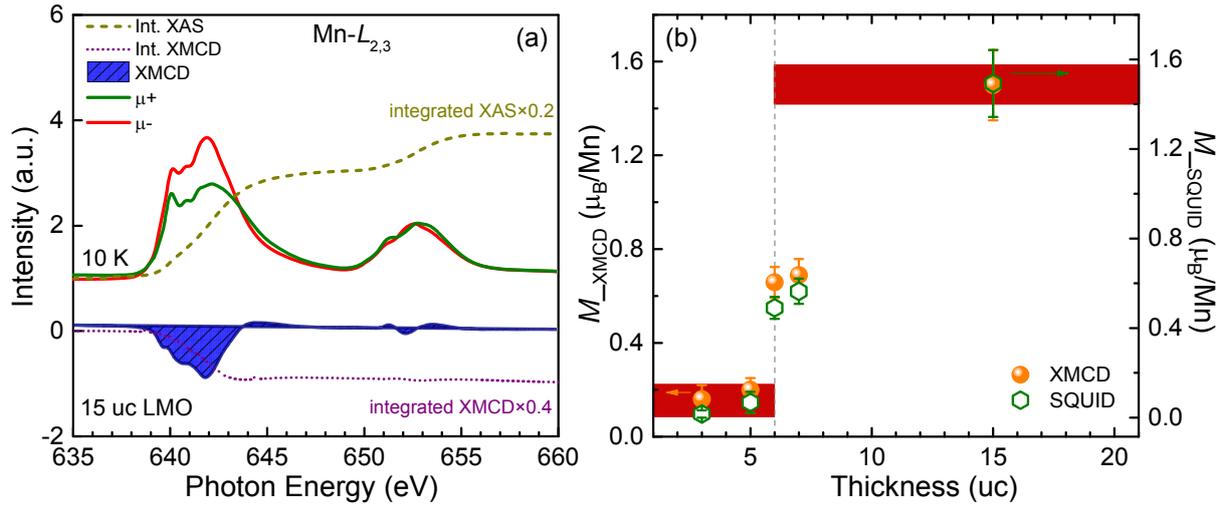

**Figure 5. Magnetic contribution of Mn ions.** (a) Typical XAS/XMCD and their integration spectra of 15 uc-LMO at Mn-$L_{2,3}$ edges at 10 K. Sum rules were used to calculate the average moment of Mn. (b) Thickness-dependent XMCD-derived and SQUID-derived magnetic moments at 10 K.

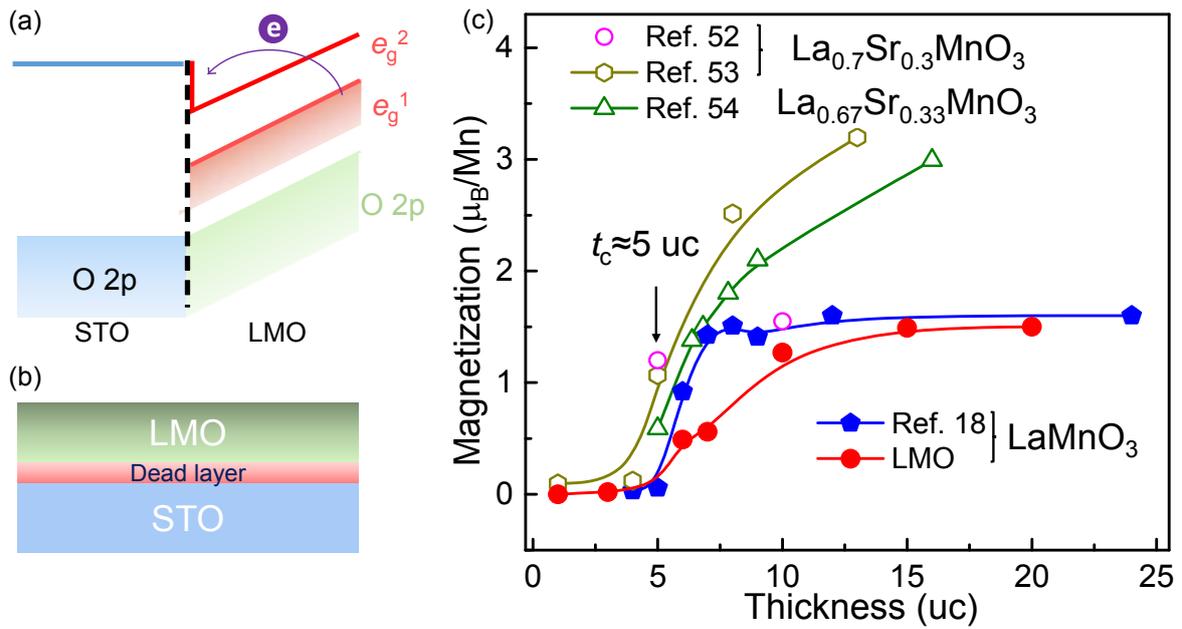

**Figure 6. Electronic reconstruction and dead-layer behavior in manganite films.** (a) The schematic of the band diagram of electronic reconstruction at the interface of LMO/STO. (b) Sketch of the dead layer at the interface of LMO/STO. (c) The thickness-dependent saturated magnetization for LMO (Ref. [18] and our data) and LSMO films (Refs. [52-54]). A critical thickness of ~2 nm (5 uc) for the suppression of magnetization in the vicinity of the interface is often observed.

# Supporting Information

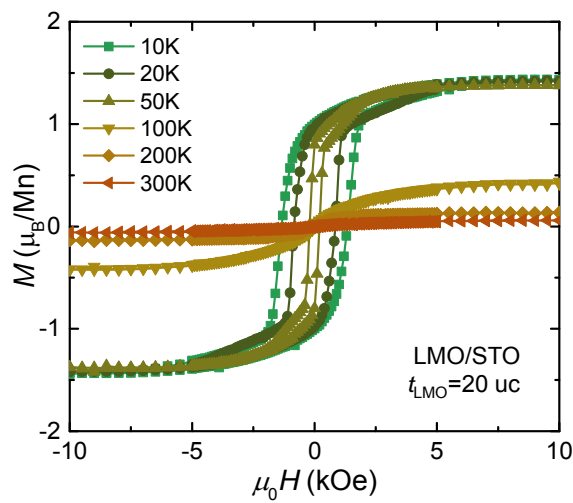

**Figure S1.** Bulk magnetic properties of the LMO films. $M(H)$ hysteresis loops for 20-uc LMO grown on STO at $1\times10^{-3}$ mbar oxygen pressure at various temperatures.

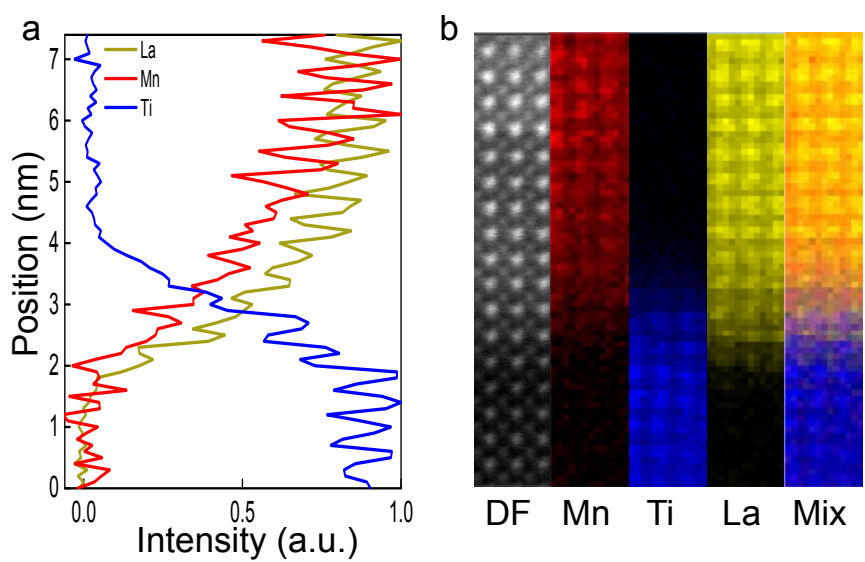

**Figure S2.** Intermixing in LMO/STO heterostructures. (a) Elemental profiles obtained from the EELS maps of the 10-uc LMO. (b) EELS elemental map with Mn in red, Ti in blue and La in yellow and the mixed image. At the interface, Mn/Ti signal can be found in a few atomic layers of LMO/STO, while La appears in the deeper regions inside the STO layer. The LMO layer is chemically wider in the La image than those in the Mn and Ti maps. An intermixing of Mn and Ti signals can be also found within a few atomic layers of LMO/STO, while La appears in the slightly deeper regions inside the STO layers. This asymmetry cation intermixing is widely observed at the interface regions between manganites and STO.[1-3]

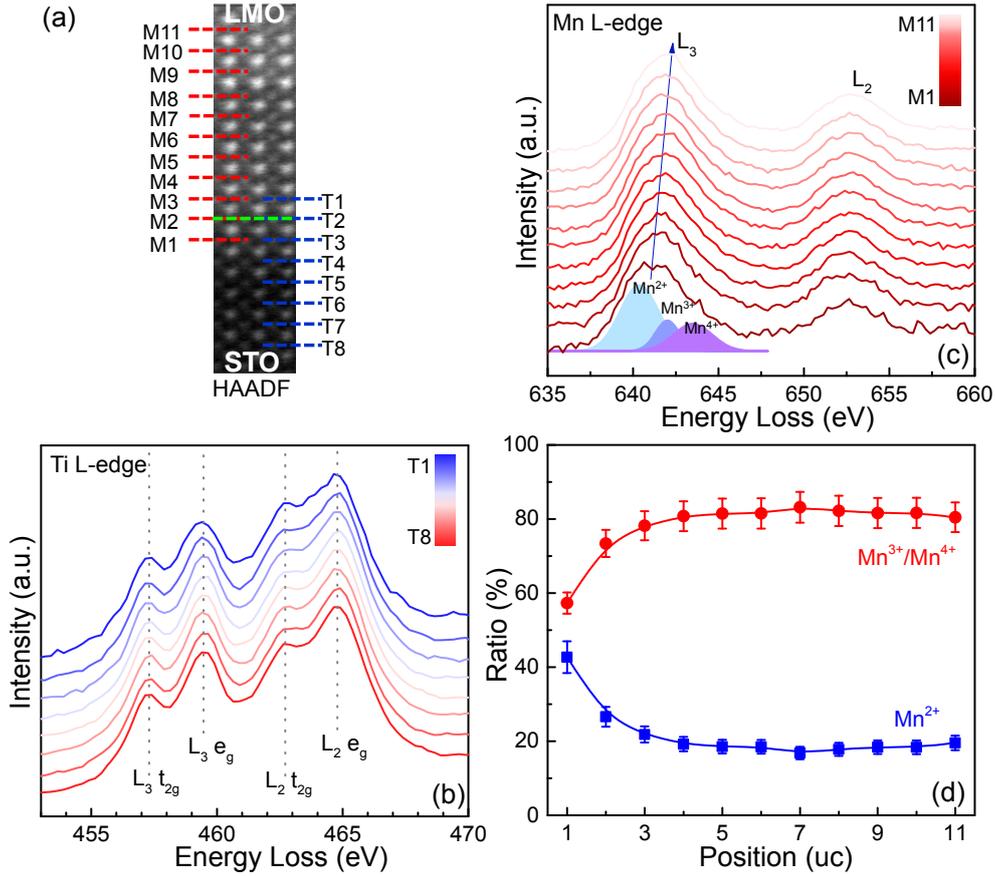

**Figure S3.** EELS analysis of Mn valence states of the LMO film grown at $P_{O2}=1\times10^{-3}$ mbar. (a) HAADF image of the corresponding area indicated by the red box in Figure 3a. The green dashed line indicates the interface between LMO and STO. (b) EELS profile of Ti-$L_{2,3}$ edges of each layer from T1 to T8. (c) EELS profile of Mn-$L_{2,3}$ edges of each layer in the selected interfacial area. (d) Different ratios of $Mn^{2+}$ and $Mn^{3+}/Mn^{4+}$ fitted from the EELS profiles in each layer.

To determine the $Mn^{3+}$ and $Mn^{4+}$ ionic content in each sample, we simultaneously and self-consistently fit our experimental data using spectra calculated for the respective Mn species via multiplet simulations. It shows that a significant fraction of $Mn^{2+}$ dominates at the interface and then remains a constant ratio. Otherwise, $Mn^{3+}/Mn^{4+}$ increases firstly and then remains unchanged with increasing the thickness of LMO.

All EELS curves of Ti-$L_{2,3}$ edges show a similar sharp multiplet structure. Remarkably, there is no peak shift in the detection limit of EELS, suggesting that there is no variation of Ti valence state. Contrary to a previous analysis of Ti edge yielding a change in the valence of $Ti^{3+}$ below the $Ti^{4+}$ state,[4, 5] the valence state of each-uc Ti in LMO/STO keeps the nominal $Ti^{4+}$ of the bulk STO even in the intermixing region of LMO/STO. The unchanged valence state of $Ti^{4+}$ in our LMO/STO heterostructures is also consistent with the previous STEM-EELS measurements of $La_{1-x}Sr_xMnO_3$/STO ($0 \leq x \leq 0.5$) heterostructures.[3]

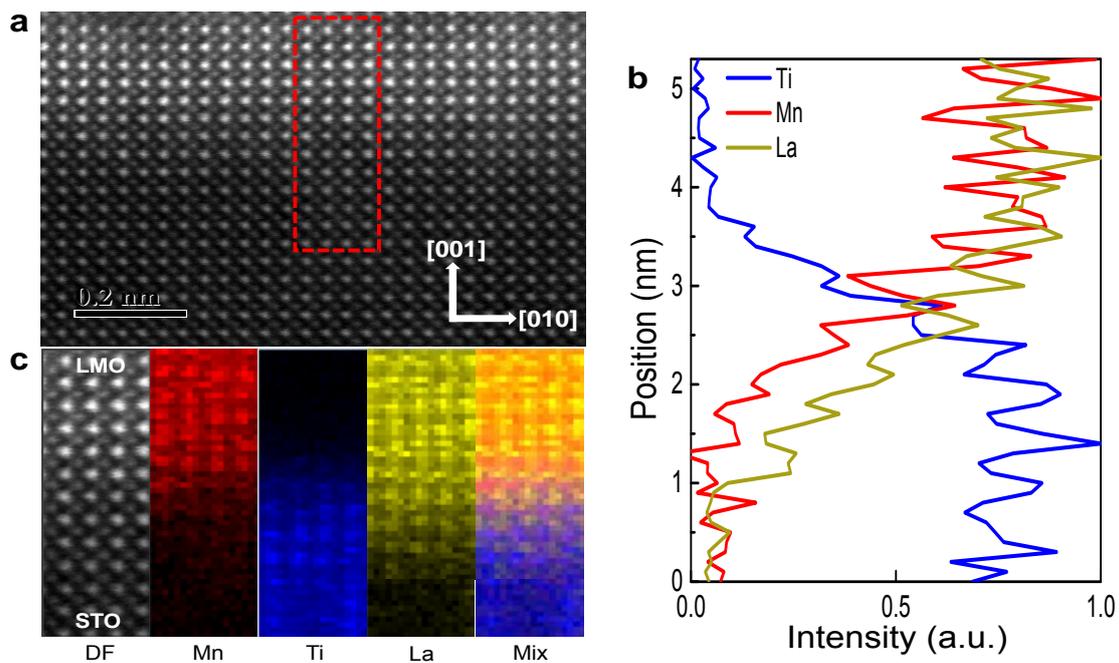

**Figure S4.** TEM characterization of 5-uc LMO/STO heterostructure. (a) Cross-sectional STEM-HAADF image. (b) Elemental profiles obtained from the EELS maps. At the interface, cation intermixing is also observed. (c) Dark-field (DF) image of the corresponding area indicated by the red box in panel **a**, and EELS elemental maps of Mn, Ti, La and the mixed image.

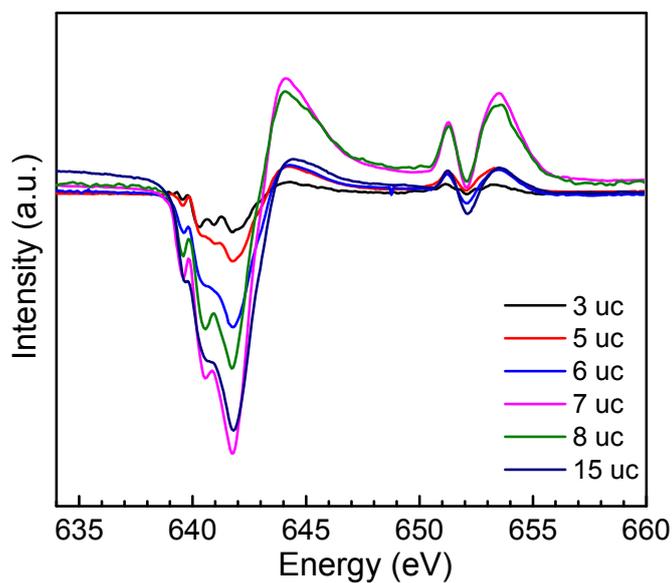

**Figure S5.** Mn XMCD spectra of the LMO/STO heterostructures with the varied LMO overlayer thickness. All LMO thin films are deposited at $P_{O2}= 1\times10^{-3}$ mbar and the spectra are obtained at $T = 10$ K.

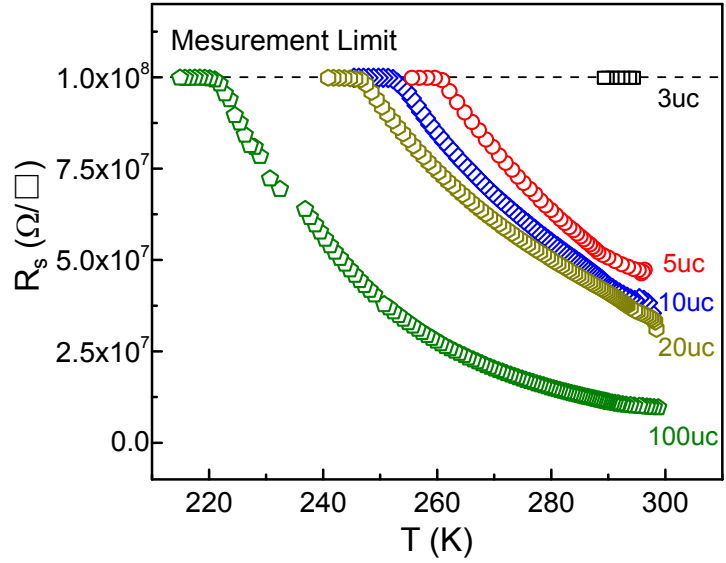

**Figure S6.** Transport properties of LMO/STO heterostructures. Sheet resistance of 3, 5, 10, 20 and 100 uc LMO on STO as a function of temperature. All LMO films exhibit insulating behaviors and beyond the measurement limit as temperature decreases. We note that, contrary to the LAO/STO system where the interface becomes metallic above the critical thickness, our film remain insulating. This is due to the difference of band gap between LAO and LMO. The large band gap of LAO leads to the electron transfer to STO.[6]

Transport properties are determined in a Van der Pauw four-probe configuration with a Quantum Design physical properties measurement system (PPMS) in the range of 2-300 K.

**Sum-rules**

The magnitude of magnetization for various-thickness LMO films is estimated quantitatively by XMCD spin sum-rule, which yields the average magnetization by using the following equations:[7]

$$m_{orb} = -\frac{4\int_{L_3+L_2}(\mu^+ - \mu^-)d\varpi}{3\int_{L_3+L_2}(\mu^+ + \mu^-)d\varpi}n_h, \qquad (1)$$

$$m_{spin} = -\frac{6\int_{L_3}(\mu^+ - \mu^-)d\varpi - 4\int_{L_3+L_2}(\mu^+ - \mu^-)d\varpi}{\int_{L_3+L_2}(\mu^+ + \mu^-)d\varpi} \times n_h(1+\frac{7<T_z>}{2<S_z>})^{-1}. \qquad (2)$$

where $\mu^+$ and $\mu^-$ are the absorption intensity with left and right circular polarized X-rays, $n_h$ is the number of holes in $d$ shells, and in this system $n_h$ is 6.[5] $<T_z>$ is the expected value of the magnetic dipole operator and $2<S_z>$ is the value of $m_{spin}$ in Hartree atomic units, which could be omitted during the calculation.[8]

**Dead-layer effect in manganite thin films**

The dead-layer behavior is widely observed in other ferromagnetic manganite films, such as $La_{0.7}Sr_{0.3}MnO_3$ (LSMO) and LMO.[9-11] The mechanism underlying this dead-layer effect remains elusive but might be due to the accumulation of $Mn^{2+}$ and/or intermixing of cations or the other external effects, which always shows a critical thickness ($t_c$) of ~2 nm (i.e., ~5 uc) for STO as the substrate.

Furthermore, LAO and LSAT are both polar oxides. Growing LMO on LAO and LSAT substrates still exhibits the clear ferromagnetism although there are no polar discontinuity. This is contradictory with the previous results:[6] 12-uc LMO grown LAO measured by scanning SQUID shows non-ferromagnetism. One possible reason without ferromagnetism can be attributed to the fact that 12 uc film is not thick enough beyond the dead-layer region of manganite grown on LAO and LSAT substrates. Notably, the dead-layer thickness is substrate-dependent. For example, the $t_c$ of the dead-layer behavior in LSMO films grown on STO, LAO and LSAT is ~8 uc, ~20 uc and ~17 uc, respectively.[10, 11] Besides, the oxygen ions in STO can diffuse over several micrometers at high temperatures during the growth.[12] This supplies more oxygen diffusing into the LMO layers, leading to the higher $M_s$ and coercivity of LMO/STO heterostructures in comparison to the case of LMO grown on LAO and LSAT substrates.

**References**
[1] Z. Yuan, J. Ruan, L. Xie, X. Pan, D. Wu, P. Wang, *Appl. Phys. Lett.* **2017**, 110, 171602.
[2] L. F. Kourkoutis, J. H. Song, H. Y. Hwang, D. A. Muller, *Proc. Natl. Acad. Sci.* **2010**, 107, 11682.
[3] J. A. Mundy, Y. Hikita, T. Hidaka, T. Yajima, T. Higuchi, H. Y. Hwang, D. A. Muller, L. F. Kourkoutis, *Nat. Commun.* **2014**, 5, 3464.
[4] J. Garcia-Barriocanal, F. Y. Bruno, A. Rivera-Calzada, Z. Sefrioui, N. M. Nemes, M. Garcia-Hernandez, J. Rubio-Zuazo, G. R. Castro, M. Varela, S. J. Pennycook, C. Leon, J. Santamaria, *Adv Mater* **2010**, 22, 627.
[5] J. Garcia-Barriocanal, J. C. Cezar, F. Y. Bruno, P. Thakur, N. B. Brookes, C. Utfeld, A. Rivera-Calzada, S. R. Giblin, J. W. Taylor, J. A. Duffy, S. B. Dugdale, T. Nakamura, K. Kodama, C. Leon, S. Okamoto, J. Santamaria, *Nat. Commun.* **2010**, 1, 82.
[6] X. R. Wang, C. J. Li, W. M. Lü, T. R. Paudel, D. P. Leusink, M. Hoek, N. Poccia, A. Vailionis, T. Venkatesan, J. M. D. Coey, E. Y. Tsymbal, Ariando, H. Hilgenkamp, *Science* **2015**, 349, 716.
[7] D. Yi, J. Liu, S. Okamoto, S. Jagannatha, Y. C. Chen, P. Yu, Y. H. Chu, E. Arenholz, R. Ramesh, *Phys Rev Lett* **2013**, 111, 127601.
[8] P. Yu, J. S. Lee, S. Okamoto, M. D. Rossell, M. Huijben, C. H. Yang, Q. He, J. X. Zhang, S. Y. Yang, M. J. Lee, Q. M. Ramasse, R. Erni, Y. H. Chu, D. A. Arena, C. C. Kao, L. W. Martin, R. Ramesh, *Phys. Rev. Lett.* **2010**, 105, 027201.
[9] R. P. Borges, W. Guichard, J. G. Lunney, J. M. D. Coey, F. Ott, *J. Appl. Phys.* **2001**, 89, 3868.
[10] M. Huijben, L. W. Martin, Y. H. Chu, M. B. Holcomb, P. Yu, G. Rijnders, D. H. A. Blank, R. Ramesh, *Phys. Rev. B* **2008**, 78, 094413.
[11] R. Peng, H. C. Xu, M. Xia, J. F. Zhao, X. Xie, D. F. Xu, B. P. Xie, D. L. Feng, *Appl. Phys. Lett.* **2014**, 104, 081606.
[12] W. Niu, Y. Gan, Y. Zhang, D. V. Christensen, M. v. Soosten, X. Wang, Y. Xu, R. Zhang, N. Pryds, Y. Chen, *Appl. Phys. Lett.* **2017**, 111, 021602.